\documentclass{article}
\begin{document}
\title{\bf Locality and the Greenberger-Horne-Zeilinger Theorem }
\author{A. Fahmi$^{1}$\footnote{ fahmi@theory.ipm.ac.ir} and M. Golshani$^{1,2}$\footnote{golshani@ihcs.ac.ir}
\vspace{7mm}\\
{\small $^{1}$Institute for Studies in Theoretical Physics and
Mathematics (IPM)}, {\small P. O. Box 19395-5531, Tehran,
Iran.}\\
{\small $^{2}$Department of Physics, Sharif University of
Technology,} {\small P. O. Box 11365-9161, Tehran, Iran.}}

\date{{\small{(\today)}}}
\maketitle

\begin{abstract}
In all local realistic theories worked out till now, locality is
considered as a basic assumption. Most people in the field
consider the inconsistency between local realistic theories and
quantum mechanics to be a result of non-local nature of quantum
mechanics. In this Paper, we derive the
Greenberger-Horne-Zeilinger type theorem for particles with
instantaneous (non-local) interactions at the hidden-variable
level. Then, we show that the previous contradiction still exists
between quantum mechanics and non-local hidden variable models.
\end{abstract}

\maketitle
\section{Introduction}
One of the main problems in physics that has attracted physicists'
attention in recent years is locality. This notion has different
meanings, interpretations and applications in different fields of
studies. Physicists consider locality principle as a physical
constraint which should be satisfied by any new theory. Quantum
mechanics (QM) has been challenging this principle for a long
period. Non-locality in QM, however, enters into calculations as a
consequence of the entanglement between some appropriate degrees
of freedom of two separated particles, which makes them to show a
correlated behavior. Entanglement has emerged as one of the most
striking feature of quantum mechanics. However, an exact
quantitative relation between non-locality and entanglement is not
been known. The non-local property of QM was first demonstrated by
Einstein-Podolsky-Rosen (EPR) \cite{EPR}, who explicitly suggested
that any physical theory must be both local and realistic. The
manifestation of these conditions was then appeared in the
so-called Bell inequality \cite{Bell}, where locality is a crucial
assumption, violated by quantum mechanical predictions. From the
so-called Bell's inequalities one can infer Bell's theorem which
states that: " \emph{In a certain experiment all realistic local
theories with hidden variables are incompatible with quantum
mechanics} ". In Bell's theorem, the locality assumption was
involved quantitatively for the first time. In the other words; If
the result of measurement on the particle $1$ $(2)$ is called $A$
$(B)$, then, in a local hidden variable (LHV) model, the locality
condition is defined as:
\begin{eqnarray}
A(a,b,\lambda)=A(a,\lambda), \hspace{.5cm}
B(a,b,\lambda)=B(b,\lambda).
\end{eqnarray}
In the above expression the result of a measurement on the
particle $1$  $(2)$ is independent of parameters and the result of
the measurement performed on the particle $2$  $(1)$. The various
experiments have been performed to distinguish between QM and
local realistic theories \cite{As}. The existing contradiction
between QM and LHV models, and also the violation of Bell's
inequality in experiments lead us to conclude that we may doubt on
one of the initial assumptions of Bell's inequality, i.e, locality
or reality. Of course, there are some people who believe that
other assumptions and loopholes might be involved, instead of the
locality and reality assumptions \cite{Barre,As1}. For example, in
a recent paper, Barrett \emph{et al.} \cite{Barre} used the
two-side memory loophole, in which the hidden variables of $n$
pair can depend on the previous measurement choices and outcomes
in both wings of the experiment. They have shown that the two-side
memory loophole allows a systematic violation of the CHSH
inequality.

Bell's inequality has been derived in different ways
\cite{CHSH,CH,KS}. Greenberger \emph{et al.} \cite{GHZ}, Hardy
\cite{Hardy} and Cabello \cite{Cab} have shown that it is possible
to demonstrate the existence of non-locality for the case of more
than two particles without using any inequality.  In another
model, Scarani and Gisin \cite{Gisin} considered some superluminal
hidden communication or influences to reproduce the non-local
correlations. There exists another attempting to clarify QM
properties. For example, Zeilinger \emph{et al.} \cite{Zei}
suggested information explanation of QM. Others works on this
subject can be summarized as follows: The extension of Bell's
inequality and Greenberger, Horne, Zeilinger (GHZ) theorem to
continuous-variables \cite{Wenger}; The Bell-type inequality that
involves all possible settings of the local measurements apparatus
\cite{Zuk}; The extension of local hidden variable models (LHV) to
multiparticle and multilevel systems \cite{Cab1}; The violation of
Bell's inequality beyond Cirelson's bound \cite{Cab2}.

On the other hand, as is known, Bell shows that it is impossible
to reproduce the Bell correlation function by using a local
realistic model. Some people extended Bell's approach, by
considering realistic interpretation of QM and showed that exact
simulation of the Bell correlation (singlet state) is possible by
using local hidden variables augmented by just one bit of
classical communication \cite{Brass,Bacon}. Hence, it has recently
been shown that all causal correlations between two parties which
respectively give one bit outputs $a$ and $b$ upon receiving one
bit inputs $x$ and $y$ can be expressed as convex combinations of
local correlations (i.e., correlations that can be simulated with
local random variables) and non-local correlations of the form
$a+b=x\cdot y$ \hspace{.2cm} \emph{mod 2}. They showed that a
single instance of the latter elementary non-local correlation
suffices to simulate exactly all possible projective measurements
that can be performed on the singlet state of two qubits, with no
communication needed at all \cite{Cerf}. Although, these theories
explain some part of QM, but they could not present a complete
description of QM. For example, in the Popescu and Rohrlich
non-local machine \cite{Cerf}, we have not all of QM properties,
it has been shown that entanglement swapping is not simulated by a
non-local machine \cite{Short} and quantum multiparties
correlations arising from measurements on a cluster state cannot
be simulated by a $n$ non-local boxes, for any $n$ \cite{BP}.

In this Paper, we derive the GHZ-type \cite{GHZ} theorem for
particles with instantaneous (non-local) interactions at the
hidden-variable level. Then, \emph{we show that the previous
contradiction still exists between QM and non-local hidden
variable models.}

\section{Locality Condition in the Greenberger-Horne-Zeilinger  Theorem}
Greenberger, Horne and Zeilinger (GHZ) showed the consequence of
Bell's theorem in a different way \cite{GHZ}, using a system which
consists of three or more correlated spin-$\frac{1}{2}$ particles.
GHZ argued that if quantum mechanical predictions hold true for
the entangled three-particle or four-particle states, then the
local hidden variable theories cannot reproduce quantum mechanical
results. The GHZ theorem is, in fact, a synthesis of Bell's
theorem \cite{Bell} and Kochen-Specker's theorem \cite{KS}, and it
indicates that we cannot attribute values to the results of
simultaneous measurements of three or more correlated particles,
without encountering a mathematical inconsistency. This theorem
provides a new test for the evaluation of concepts like locality
and non-contextuality on the basis of complete quantum
correlations (for multi-particle entangled states, the assumption
of non-contextuality is usually taken to be equivalent to
locality). GHZ considerd a system of four spin $\frac{1}{2}$
particles so that particles $1$ and $2$ move freely in the
positive $z$-direction and particles $3$ and $4$ in the negative
$z$-direction. The Stern-Gerlach orientation analyzers are
$\widehat{n}_{1},\widehat{n}_{2},\widehat{n}_{3}$ and
$\widehat{n}_{4}$ for the beams of particles $1,2,3$ and $4$,
respectively. If these four particles result from the decay of a
single spin-1 particle into a pair of spin-1 particles, each of
which decays into a pair of spin-$\frac{1}{2}$ particles, then,
with the $z$ component of spin initially zero and remaining so
throughout the decay process, the quantum mechanical spin state of
the four particles is:
\begin{eqnarray*}
|\psi\rangle=\frac{1}{\sqrt{2}}[|+\rangle_{1}|+\rangle_{2}|-\rangle_{3}|-\rangle_{4}-
|-\rangle_{1}|-\rangle_{2}|+\rangle_{3}|+\rangle_{4}].
\end{eqnarray*}
The expectation value
$E(\widehat{n}_{1},\widehat{n}_{2},\widehat{n}_{3}
\widehat{n}_{4})$ of the product of outcomes, when the
orientations are as indicated, is:
\begin{eqnarray}
E(\widehat{n}_{1},\widehat{n}_{2},\widehat{n}_{3}
\widehat{n}_{4})&=&\langle\psi|(\overrightarrow{\sigma}_{1}.\widehat{n}_{1})
(\overrightarrow{\sigma}_{2}.\widehat{n}_{2})
(\overrightarrow{\sigma}_{3}.\widehat{n}_{3})
(\overrightarrow{\sigma}_{4}.\widehat{n}_{4}) |\psi\rangle
=\cos\theta_{1}\cos\theta_{2}\cos\theta_{3}\cos\theta_{4}\\\nonumber
&&-\sin\theta_{1}\sin\theta_{2}\sin\theta_{3}\sin\theta_{4}\times
\cos(\phi_{1}+\phi_{2}-\phi_{3}-\phi_{4})
\end{eqnarray}
where $(\theta_{i}\phi_{i})$ are polar and azimuthal angles of
$\widehat{n}_{i}$. For simplicity, we shall restrict our attention
to $\widehat{n}$'s in the $x-y$ plane, so that:
\begin{eqnarray}\label{3}
E(\widehat{n}_{1},\widehat{n}_{2},\widehat{n}_{3}
\widehat{n}_{4})=E(\phi_{1},\phi_{2},\phi_{3},\phi_{4})=-
\cos(\phi_{1}+\phi_{2}-\phi_{3}-\phi_{4})
\end{eqnarray}
EPR's assumptions in the GHZ argument can be adapted to the
four-particle situation as follows:\\
\emph{(i) Perfect correlation}: Knowledge of the outcomes for any
three particles enables a prediction with certainty for the
outcome
of the fourth.\\
\emph{(ii) Locality}: Since the four particles are arbitrarily far
apart at the time of measurement, and are assumed not to interact,
hence no real change can takes place in any one of them as a
consequence of what is done on the other three.
\\\emph{(iii) Reality}: If, without in any way disturbing a
system, we can predict with certainty (i.e. with probability equal
to unity ) the value of a physical quantity, then there exists an
element of physical reality corresponding to this physical
quantity.\\
\emph{(iv) Completeness}: Every element of the physical reality
must have a counterpart in the physical theory.\\
Noting the above premises, GHZ defined four functions
$A(\phi_{1},\lambda),B(\phi_{2},\lambda),C(\phi_{3},\lambda),
D(\phi_{4},\lambda)$ with values $+1$ and $-1$, these functions
being the outcomes of spin measurement along the respective
directions when the complete state of the four particles is
specified by $\lambda$. Using the above premises and some algebra,
GHZ could derive following relations \cite{GHZ}:
\begin{eqnarray*}
A(2\phi,\lambda)=A( 0 ,\lambda)= Const., \hspace{1cm}\forall
\hspace{.2cm} \phi
\end{eqnarray*}
\begin{eqnarray}
A(\phi'+\pi,\lambda)=-A( 0 ,\lambda)=Const., \hspace{1cm}\forall
\hspace{.2cm}  \phi'
\end{eqnarray}
GHZ showed that the above relations are inconsistent with each
other for $\phi=\frac{\pi}{2}, \phi'=0$. Thus, they come to the
conclusion that a hidden inconsistency is present in premises
\emph{(i)-(iv)}. In our approach to the GHZ theorem, we keep the
hypotheses \emph{i,iii,iv} as before and replace \emph{ii}
by the following statement \cite{Fa}:\\
\emph{(ii') Non-locality:} at the time of measurement, when the
four particles interact, a real change can take place in a
particle as a consequence of any thing that may be done on the
other three particles.\\
The above condition changes our view about locality and the GHZ
theorem. In \cite{Fa}, we defined non-locality condition at the
hidden variable level as:
\begin{eqnarray}\label{5}
A(\phi_{1},\phi_{2},\phi_{3},\phi_{4},\lambda)=f_{A}(\phi_{1},\lambda)
g_{A}(\phi_{2},\lambda)h_{A}(\phi_{3},\lambda)k_{A}(\phi_{4},\lambda)\nonumber\\
\end{eqnarray}
Similar relations hold for $B, C$ and $D$. Using the above form of
non-locality, we have:
\begin{eqnarray*}
A(2\phi_{1},2\phi_{2},2\phi_{3},2\phi_{4},\lambda)=
A(0,0,0,0,\lambda)
\end{eqnarray*}
\begin{eqnarray*}
A(\theta_{1}+\pi,\theta_{2}+\pi,\theta_{3}+\pi,\theta_{4},\lambda)=
-A(\theta_{1},\theta_{2},\theta_{3},\theta_{4},\lambda)
\end{eqnarray*}
For
$\overrightarrow{\phi}=(\frac{\pi}{2},\frac{\pi}{2},\frac{\pi}{2},0)$
and $\overrightarrow{\theta}=(0,0,0,0)$ one gets:
\begin{eqnarray*}
A(\pi,\pi,\pi,0,\lambda)= A(0,0,0,0,\lambda)
\end{eqnarray*}
\begin{eqnarray*}
A(\pi,\pi,\pi,0,\lambda)= -A(0,0,0,0,\lambda)
\end{eqnarray*}
which leads to the usual GHZ result.
\section{Extension of Non-locality Condition to the Greenberger-Horne-Zeilinger  Theorem}
In this section we would like to extend our non-locality condition
to a more general form. This form can be applied to a very large
class of non-local hidden variable theories. One generalization of
eq. (\ref{5}) is in the following manner:
\begin{eqnarray}\label{6}
A(\phi_{1},\phi_{2},\phi_{3},\phi_{4},\lambda)
=\sum_{j_{A}}f_{A}^{j_{A}}(\phi_{1},\lambda)
g_{A}^{j_{A}}(\phi_{2},\lambda)h_{A}^{j_{A}}(\phi_{3},\lambda)k_{A}^{j_{A}}(\phi_{4},\lambda)
\end{eqnarray}
Where $f, g, h$ and $k$ are arbitrary functions that depend on
their arguments with:
\begin{eqnarray*}
-1\leq A(\phi_{1},\phi_{2},\phi_{3},\phi_{4},\lambda)\leq1
\end{eqnarray*}
and similar relations hold for $B,C,D$. This type of
generalization is not special. Actually, for any realistic
variable we can make such an assumption. Also, these variables
could not be considered as a mathematical generalization but as
something that could convey some
physical meaning \cite{Bohm1}.  \\

The product of these physical variables is:
\begin{eqnarray}\label{7}
ABCD=\prod_{i=A}^{D}\sum_{j_{i}}f_{i}^{j_{i}}(\phi_{1},\lambda)
g_{i}^{j_{i}}(\phi_{2},\lambda)h_{i}^{j_{i}}(\phi_{3},\lambda)k_{i}^{j_{i}}(\phi_{4},\lambda)
\end{eqnarray}
We define the correlation function $E(\phi_{1}, \phi_{2},
\phi_{3}, \phi_{4})$ in the form:
\begin{eqnarray}\label{77}
E(\phi_{1}, \phi_{2}, \phi_{3}, \phi_{4})=\int
\prod_{i=A}^{D}\sum_{j_{i}}f_{i}^{j_{i}}(\phi_{1},\lambda)
g_{i}^{j_{i}}(\phi_{2},\lambda)h_{i}^{j_{i}}(\phi_{3},\lambda)k_{i}^{j_{i}}(\phi_{4},\lambda)
\rho(\lambda)d\lambda
\end{eqnarray}
where the probability distribution function for the uncontrollable
hidden variable $\lambda$ is represented by $\rho(\lambda)$, with:
\begin{eqnarray*}
\int \rho(\lambda)d\lambda=1, \hspace{1cm} \rho(\lambda)\geq1
\end{eqnarray*}
Using the relations (\ref{3}) for the expectation value of the
spin correlation, one can show that for
$\phi_{1}+\phi_{2}-\phi_{3}-\phi_{4}=0$:
\begin{eqnarray}\label{8}
\prod_{i=A}^{D}\sum_{j_{i}}f_{i}^{j_{i}}(\phi,\lambda)
g_{i}^{j_{i}}(\phi,\lambda)h_{i}^{j_{i}}(\phi,\lambda)k_{i}^{j_{i}}(\phi,\lambda)=-1
\end{eqnarray}
\begin{eqnarray}\label{9}
\prod_{i=A}^{D}\sum_{j_{i}}f_{i}^{j_{i}}(\phi,\lambda)
g_{i}^{j_{i}}(\phi',\lambda)h_{i}^{j_{i}}(\phi,\lambda)k_{i}^{j_{i}}(\phi',\lambda)=-1
\end{eqnarray}
\begin{eqnarray}\label{10}
\prod_{i=A}^{D}\sum_{j_{i}}f_{i}^{j_{i}}(\phi,\lambda)
g_{i}^{j_{i}}(\phi',\lambda)h_{i}^{j_{i}}(\phi',\lambda)k_{i}^{j_{i}}(\phi,\lambda)=-1
\end{eqnarray}
\begin{eqnarray}\label{11}
\prod_{i=A}^{D}\sum_{j_{i}}f_{i}^{j_{i}}(\phi',\lambda)
g_{i}^{j_{i}}(\phi,\lambda)h_{i}^{j_{i}}(\phi/2+\phi'/2,\lambda)k_{i}^{j_{i}}(\phi/2+\phi'/2,\lambda)=-1
\end{eqnarray}

By changing $\phi'$ in Eq. (\ref{9}), we get a set of equations
(with constant $\phi$) which can be written as a matrix equation:\\
\begin{eqnarray*}
\left( \begin{array}{cccccccc}
\prod_{i=A}^{D} g_{i}^{1}(\phi_{1}')k_{i}^{1}(\phi_{1}') & . & . &  & \prod_{i=A}^{D} g_{i}^{j_{i}}(\phi_{1}')k_{i}^{j_{i}}(\phi_{1}') & . & . &\prod_{i=A}^{D} g_{i}^{n}(\phi_{1}')k_{i}^{n}(\phi_{1}')  \\
\prod_{i=A}^{D} g_{i}^{1}(\phi_{2}')k_{i}^{1}(\phi_{2}') & . & . &  & \prod_{i=A}^{D} g_{i}^{j_{i}}(\phi_{2}')k_{i}^{j_{i}}(\phi_{2}') & . & . &\prod_{i=A}^{D} g_{i}^{n}(\phi_{2}')k_{i}^{n}(\phi_{2}')  \\
            .                 & . & . &  &             .                  & . & . &            .                    \\
            .                 & . & . &  &             .                  & . & . &            .                    \\
\prod_{i=A}^{D} g_{i}^{1}(\phi_{k}')k_{i}^{1}(\phi_{k}') & . & . &  & \prod_{i=A}^{D} g_{i}^{j_{i}}(\phi_{k}')k_{i}^{j_{i}}(\phi_{k}') & . & . &\prod_{i=A}^{D} g_{i}^{n}(\phi_{k}')k_{i}^{n}(\phi_{k}')  \\
            .                 & . & . &  &             .                  & . & . &            .                    \\
            .                 & . & . &  &             .                  & . & . &            .                    \\
\prod_{i=A}^{D} g_{i}^{1}(\phi_{n}')k_{i}^{1}(\phi_{n}') & . & . &  & \prod_{i=A}^{D} g_{i}^{j_{i}}(\phi_{n}')k_{i}^{j_{i}}(\phi_{n}') & . & . &\prod_{i=A}^{D} g_{i}^{n}(\phi_{n}')k_{i}^{n}(\phi_{n}')  \\

\end{array}
\right)
\end{eqnarray*}
\begin{eqnarray}\label{12}
\times \left(
\begin{array}{c}
\prod_{i=A}^{D} f_{i}^{1}(\phi_{1})h_{i}^{1}(\phi_{1}) \\
\prod_{i=A}^{D} f_{i}^{2}(\phi_{1})h_{i}^{2}(\phi_{1}) \\
                  .                 \\
                  .                 \\
\prod_{i=A}^{D} f_{i}^{j_{i}}(\phi_{1})h_{i}^{j_{i}}(\phi_{1}) \\
                  .                 \\
                  .                 \\
\prod_{i=A}^{D} f_{i}^{n}(\phi_{1})h_{i}^{n}(\phi_{1}) \\
\end{array}
\right)=-\left(
\begin{array}{c}
 1 \\
 1 \\
 . \\
 . \\
 1 \\
 . \\
 . \\
 1 \\
\end{array}
\right)
\end{eqnarray}
In the above matrix equation we have droped some of unnecessary
indices for simplicity and $\phi'$'s are arbitrary angles (see
appendix B).

Also, we can get similar matrix equations by changing $\phi$ and
holding $\phi'$ constant. By calculating
$\prod_{i=A}^{D}f_{i}^{j_{i}}(\phi_{l},\lambda)
h_{i}^{j_{i}}(\phi_{l},\lambda) $ from each of these equations, we
have:
\begin{eqnarray}\label{13}
\prod_{i=A}^{D}f_{i}^{j_{i}}(\phi_{l},\lambda)
h_{i}^{j_{i}}(\phi_{l},\lambda)=
\prod_{i=A}^{D}f_{i}^{j_{i}}(\phi_{m},\lambda)
h_{i}^{j_{i}}(\phi_{m},\lambda)\nonumber\\
\end{eqnarray}
In above relation, $f_{i}^{j_{i}}$ is the $j$th term of Eq.
(\ref{6}) related to $i$th party. (To distinguish between the
functions of different parties we have tagged them, for example,
by $j_{i}$). Similarly, by using Eq. (\ref{10}), we have:
\begin{eqnarray}\label{14}
\prod_{i=A}^{D}f_{i}^{j_{i}}(\phi_{l},\lambda)
k_{i}^{j_{i}}(\phi_{l},\lambda)=
\prod_{i=A}^{D}f_{i}^{j_{i}}(\phi_{m},\lambda)
k_{i}^{j_{i}}(\phi_{m},\lambda)\nonumber\\
\end{eqnarray}
A consequence of these is that:
\begin{eqnarray}\label{15}
&&\prod_{i=A}^{D}
h_{i}^{j_{i}}(\phi_{l},\lambda)(k_{i}^{j_{i}}(\phi_{l},\lambda)^{-1}=
\prod_{i=A}^{D}
h_{i}^{j_{i}}(\phi_{m},\lambda)(k_{i}^{j_{i}}(\phi_{m},\lambda))^{-1}
\end{eqnarray}
Now, we can use two approaches. First, by assuming that any term
in the Eq. (\ref{6}) is $f_{i}^{j_{i}}, g_{i}^{j_{i}},
h_{i}^{j_{i}}, k_{i}^{j_{i}}=\pm1$, Eq. (\ref{15}) can be written
in the following form (cf \cite{Fa}),
\begin{eqnarray}\label{16}
\prod_{i=A}^{D}
h_{i}^{j_{i}}(\phi_{l},\lambda)k_{i}^{j_{i}}(\phi_{l},\lambda)=
\prod_{i=A}^{D}
h_{i}^{j_{i}}(\phi_{m},\lambda)k_{i}^{j_{i}}(\phi_{m},\lambda)
\end{eqnarray}
Second, by considering that $A^{2}(\phi_{1}, \phi_{2},
\phi_{3},\phi_{4}, \lambda)= 1$ and using Eq. (\ref{6}), we can
construct matrix equations similar to (\ref{12}), which finally
lead to:
\begin{eqnarray}\label{17}
[f_{A}^{j_{A}}(\phi_{1},\lambda)]^{2}=X(\Phi, \lambda)
\end{eqnarray}
where $\Phi$ does not contain $\phi_{1}$ (see appendix A).
Although, the above equation is surprising, nevertheless, if
$f_{i}^{j_{i}}=\pm 1$, we can convince ourself with this result.
The above relation holds for any $\phi_{1}$ and similar relations
holds for $g,h,k$. If we consider $k_{i}^{j_{i}}$ type of Eq.
(\ref{17}) and apply to Eq. (\ref{15}), we get Eq. (\ref{16})
again. Now, by using Eqs. (\ref{8}), (\ref{11}) and (\ref{16}),
and after some algebra, we have:
\begin{eqnarray}\label{18}
\prod_{i=A}^{D} f_{i}^{j_{i}}(\phi_{l},\lambda) =\prod_{i=A}^{D}
f_{i}^{j_{i}}(\phi_{m},\lambda)
\end{eqnarray}
Similar relations can be derived for $g, h$ and $k$. This equation
is a quite surprising result, we would expect the change of $f, g,
h$ and $k$'s arguments, which their combination represents an
intrinsic spin quantity, to have some different outcomes.

We can repeat the whole calculation by assuming that
$\phi_{1}+\phi_{2}-\phi_{3}-\phi_{4}=\pi$ in the Eq. (\ref{77}).
Then we have:

\begin{eqnarray}\label{19}
\prod_{i=A}^{D}\sum_{j_{i}}f_{i}^{j_{i}}(\phi'+\pi,\lambda)
g_{i}^{j_{i}}(\phi,\lambda)h_{i}^{j_{i}}(\phi/2+\phi'/2,\lambda)
k_{i}^{j_{i}}(\phi/2+\phi'/2,\lambda)=1
\end{eqnarray}
From the Eqs. (\ref{8}), (\ref{16}), and (\ref{19}), we have:
\begin{eqnarray}\label{20}
\prod_{i=A}^{D}f_{i}^{j_{i}}(\phi_{s},\lambda)=-
\prod_{i=A}^{D}f_{i}^{j_{i}}(\phi_{k}+\pi,\lambda)
\end{eqnarray}
Now, considering Eqs. (\ref{18}) and (\ref{20}), it is easily seen
that for $\phi_{l}=\phi_{s}=\phi_{k}=0, \phi_{m}=\pi$, and we get:
\begin{eqnarray}\label{21}
\prod_{i=A}^{D} f_{i}^{j_{i}}(0,\lambda)&=&\prod_{i=A}^{D}
f_{i}^{j_{i}}(\pi,\lambda)\nonumber\\
\prod_{i=A}^{D}f_{i}^{j_{i}}(0,\lambda)&=&-
\prod_{i=A}^{D}f_{i}^{j_{i}}(\pi,\lambda)
\end{eqnarray}
We thus have brought to surface an inconsistency hidden in the
premisses ($i,ii', iii, iv$). We can also consider equations
(\ref{18}) and (\ref{20}) for $g, h, k$. One gets:
\begin{eqnarray}\label{22}
A(\phi_{1},\phi_{2},\phi_{3},\phi_{4},\lambda)=
A(\phi_{1}',\phi_{2}',\phi_{3}',\phi_{4}',\lambda)
\end{eqnarray}
\begin{eqnarray}\label{23}
A(\varphi_{1}+\pi,\varphi_{2},\varphi_{3},\varphi_{4},\lambda)
=-A(\varphi_{1}',\varphi_{2}',\varphi_{3}',\varphi_{4}',\lambda)
\end{eqnarray}
By using suitable angles, a discrepancy in one $A$'s would result.
So we reach the same results as the GHZ theorem, even though we
are using a special case of non-locality. Our approach rules out a
broader class of hidden variable theories.
\section{Conclusion}
Bell's theorem \cite{Bell} states that any local realistic view of
the world is incompatible with QM. This is often interpreted as
demonstrating the existence of non-locality in QM \cite{Bohm}.

In this paper, we have replaced Bell's locality condition by a
more general condition to obtain the GHZ theorem and have shown
that the same incompatibility exist for the case of non-local
realistic models. Thus, we can conclude that the disagreement in
the GHZ theorem is not necessarily due to the violation of Bell's
locality condition. Consequently, we should focus on other GHZ
assumptions to find the origin of inconsistency. Also, it is
worthy to note that one could even go further and extend the above
non-local approach to other relevant theorems such as
Kochen-Specker \cite{KS}, CH \cite{CH} and Hardy \cite{Hardy1}
theorems.

There exists an important question: do some non-local hidden
variable theories exist that our approach can be applied to them?
One can expects more general
cases of non-local theories which are incompatible with quantum
mechanics. However, it is an open problem whether one can
construct a hidden variable model, satisfying the above
requirements, to show the consistency of our argument in a more
concrete manner. Further more, in another paper, we showed that it
is not possible to reproduce a QM correlation, using only local
measurements (done by two space-like separated parties), augmented
with a classical communication having one bit information
\cite{Bacon} or by a single use of non-local box \cite{Cerf}.
Although, other people have shown that all of QM properties cannot
be simulated by a nonlocal box completely \cite{Short,BP}.

Therefore, it can be concluded that some other alternative view
points might be involved. On the other hand, these are people who
still believe that QM is a local theory \cite{Peres}, and some
others consider information as the root of the interpretation of
QM \cite{Zei}. Hence, G. Brassard and co-workers suggested the
field of communication complexity, which concerns the amount of
information that must be transmitted between two parties to
compute some function of private inputs that they hold
\cite{Brass1}. Anyway, the above arguments indicate that we must
have a deeper understanding of the notions of locality, reality
and entanglement.
\\\\
{\bf Acknowledgment}: We would like to thank P. H. Eberhard for
his comments and A. T. Rezakhani for critical reading of the
manuscript. (This work was supported under the project:
\emph{AZAD}).
\section{Appendix A: Measurement Results Matrix Equation}
In this appendix, we would like to derive the equation (\ref{17}).
As we mentioned, by using equations $A^{2}(\phi_{1}, \phi_{2},
\phi_{3},\phi_{4}, \lambda)= 1$ and (\ref{6}), we construct matrix
equation as follows:
\begin{eqnarray*}
\left( \begin{array}{cccccccc}
[g_{A}^{1}(\phi_{2}^{1})h_{A}^{1}(\phi_{3}^{1})k_{A}^{1}(\phi_{4}^{1})]^{2}
& . & . &  &
g_{A}^{j_{A}}(\phi_{2}^{1})h_{A}^{j_{A}}(\phi_{3}^{1})k_{A}^{j_{A}}(\phi_{4}^{1})g_{A}^{s_{A}}(\phi_{2}^{1})h_{A}^{s_{A}}(\phi_{3}^{1})k_{A}^{s_{A}}(\phi_{4}^{1})
& . & .
&[g_{A}^{n}(\phi_{2}^{1})h_{A}^{n}(\phi_{3}^{1})k_{A}^{n}(\phi_{4}^{1})]^{2}
\\\nonumber
[g_{A}^{1}(\phi_{2}^{2})h_{A}^{1}(\phi_{3}^{2})k_{A}^{1}(\phi_{4}^{2})]^{2}
& . & . &  &
g_{A}^{j_{A}}(\phi_{2}^{2})h_{A}^{j_{A}}(\phi_{3}^{2})k_{A}^{j_{A}}(\phi_{4}^{2})g_{A}^{s_{A}}(\phi_{2}^{2})h_{A}^{s_{A}}(\phi_{3}^{2})k_{A}^{s_{A}}(\phi_{4}^{2})
& . & . &
[g_{A}^{n}(\phi_{2}^{2})h_{A}^{n}(\phi_{3}^{2})k_{A}^{n}(\phi_{4}^{2})]^{2}  \\
            .                 & . & . &  &             .                  & . & . &            .
            \\\nonumber
            .                 & . & . &  &             .                  & . & . &            .
            \\\nonumber
[g_{A}^{1}(\phi_{2}^{k})h_{A}^{1}(\phi_{3}^{k})k_{A}^{1}(\phi_{4}^{k})]^{2}
& . & . &  &
g_{A}^{j_{A}}(\phi_{2}^{k})h_{A}^{j_{A}}(\phi_{3}^{k})k_{A}^{j_{A}}(\phi_{4}^{k})g_{A}^{s_{A}}(\phi_{2}^{k})h_{A}^{s_{A}}(\phi_{3}^{k})k_{A}^{s_{A}}(\phi_{4}^{k})
& . & .
&[g_{A}^{n}(\phi_{2}^{k})h_{A}^{n}(\phi_{3}^{k})k_{A}^{n}(\phi_{4}^{k})]^{2}
\\\nonumber
            .                 & . & . &  &             .                  & . & . &            .
            \\\nonumber
            .                 & . & . &  &             .                  & . & . &            .
            \\\nonumber
[g_{A}^{1}(\phi_{2}^{n})h_{A}^{1}(\phi_{3}^{n})k_{A}^{1}(\phi_{4}^{n})]^{2}
& . & . &  &
g_{A}^{j_{A}}(\phi_{2}^{n})h_{A}^{j_{A}}(\phi_{3}^{n})k_{A}^{j_{A}}(\phi_{4}^{n})g_{A}^{s_{A}}(\phi_{2}^{n})h_{A}^{s_{A}}(\phi_{3}^{n})k_{A}^{s_{A}}(\phi_{4}^{n})
& . & . &
[g_{A}^{n}(\phi_{2}^{n})h_{A}^{n}(\phi_{3}^{n})k_{A}^{n}(\phi_{4}^{n})]^{2}
\\\nonumber
\end{array}
\right)
\end{eqnarray*}
\begin{eqnarray}\label{23}
\times\left(
\begin{array}{c}
[f_{A}^{1}(\phi_{1})]^{2} \\
f_{A}^{1}(\phi_{1})f_{A}^{2}(\phi_{1}) \\
                  .                 \\
                  .                 \\
f_{A}^{j_{A}}(\phi_{1})f_{A}^{s_{A}}(\phi_{1}) \\
                  .                 \\
                  .                 \\\nonumber
[f_{A}^{n}(\phi_{1})]^{2} \\
\end{array}
\right)=\left(
\begin{array}{c}
 1 \\
 1 \\
 . \\
 . \\
 1 \\
 . \\
 . \\
 1 \\
\end{array}
\right)\\
\end{eqnarray}
In the above matrix equation we have dropped some of unnecessary
indices for simplicity and $\phi_{l}^{k}$'s are arbitrary angles.

We would like to solve the above matrix equation to derive column
matrix element functions $f_{A}^{j_{A}}(\phi_{1})$ with
$j_{A}=1,...,n $. It is not complicated to show that the above
matrix equation is solvable if matrix equations which have been
constructed by $A(\phi_{1},\phi_{k}, \phi_{l},\phi_{m}, \lambda)=
\pm 1$ (similar to the matrix equation with fix $\phi_{1}$ and
arbitrary $\phi^{k}_{2},\phi^{l}_{3}$ and $\phi^{m}_{4}$ with
$k=1,...,n$) have inverse. That matrix equation has the following
form:
\begin{eqnarray}\label{24}
(ghk)_{n\times n}(f)_{n\times 1}=(\pm1)_{n\times1}
\end{eqnarray}
If one of rows or columns of $(ghk)$ would be a linear combination
of other rows or columns respectively, then, $(ghk)$ matrix has
not inverse.

In the first case, where (one of rows would be equal to a linear
combination of other rows), we get to the appropriate relations:
\begin{eqnarray}\label{25}
g_{A}^{j_{A}}(\phi^{l}_{2},\lambda)=g_{A}^{j_{A}}(\phi^{q}_{2},\lambda)\nonumber\\
h_{A}^{j_{A}}(\phi^{l}_{3},\lambda)=h_{A}^{j_{A}}(\phi^{q}_{3},\lambda)\nonumber\\
k_{A}^{j_{A}}(\phi^{l}_{4},\lambda)=k_{A}^{j_{A}}(\phi^{q}_{4},\lambda)
\end{eqnarray}
In other words, $g,h,k$ are independent of their arguments. In
other cases (one of columns would be equal to linear combination
of other columns), we would have, for example:
\begin{eqnarray*}
g_{A}^{1}(\phi_{l},\lambda)h_{A}^{1}(\phi_{q},\lambda)k_{A}^{1}(\phi_{s},\lambda)=g_{A}^{2}(\phi_{l},\lambda)h_{A}^{2}(\phi_{q},\lambda)k_{A}^{2}(\phi_{s},\lambda)
\end{eqnarray*}
At lest two sentences in the $A$ expansion (\ref{6}) are equal to
each other. It is not complicated to show that after a
rearrangement of $A$ sentences, we get the same matrix equation as
(\ref{24}). After, solving the matrix equation (\ref{23}), we
would be get an appropriate result:
\begin{eqnarray}\label{26}
[f_{A}^{j_{A}}(\phi_{1},\lambda)]^{2}=X(\Phi, \lambda)\hspace{1
cm}
f_{A}^{j_{A}}(\phi_{1},\lambda)f_{A}^{s_{A}}(\phi_{1},\lambda)=Y(\Phi,
\lambda)
\end{eqnarray}
where $\Phi$ does not contain $\phi_{1}$. If we repeat equation
(\ref{23}), with the change of $\phi_{1}$ ($g,h$ and $k$ arguments
don't change), we derive similar equations as (\ref{26}), thus:
\begin{eqnarray}\label{27}
[f_{A}^{k}(\phi^{l}_{1})]^{2}=[f_{A}^{k}(\phi^{m}_{1})]^{2}\hspace{1cm}
f_{A}^{j_{A}}(\phi^{l}_{1})f_{A}^{s_{A}}(\phi^{l}_{1})=f_{A}^{j_{A}}(\phi^{m}_{1})f_{A}^{s_{A}}(\phi^{m}_{1})
\end{eqnarray}
\section{Appendix B: Correlation Function's Matrix Equation}
Matrix equation (\ref{12}) has the following form:
\begin{eqnarray}
(GK)_{n\times n}(FH)_{n\times 1}=-(1)_{n\times1}
\end{eqnarray}
We would like to solve the above matrix equation to derive column
matrix element functions $(FH)_{i,1}$, if one of rows or columns
of $(GK)$ would be a linear combination of other rows or columns
respectively, then, $(GK)$ matrix has no inverse. In the first
case (one of rows would be equal to a linear combination of other
rows), we get the appropriate relation
\begin{eqnarray}
\prod_{i=A}^{D}
g_{i}^{j_{i}}(\phi_{l}',\lambda)k_{i}^{j_{i}}(\phi_{l}',\lambda)=\prod_{i=A}^{D}
g_{i}^{j_{i}}(\phi_{k}',\lambda)k_{i}^{j_{i}}(\phi_{k}',\lambda),
\end{eqnarray}
This says that these combinations of $g$ and $k$ are independent
of their arguments ($\phi_{l}'$). In the other case (one of
columns would be equal to a linear combination of other columns),
we would have, for example,
$g_{A}^{1}(\phi_{1}',\lambda)k_{A}^{1}(\phi_{1}',\lambda)=g_{A}^{2}(\phi_{1}',\lambda)k_{A}^{2}(\phi_{1}',\lambda)$
(at lest two sentences in the $A$ expansion (\ref{6}) are equal to
each other), it is not complicated to show that after a
rearrangement of $A$ sentence and solving a matrix equation,
similar to Eq. (12), we get:
\begin{eqnarray}
f_{A}^{1}(\phi_{l},\lambda)
h_{A}^{1}(\phi_{l},\lambda)+f_{A}^{2}(\phi_{l},\lambda)
h_{A}^{2}(\phi_{l},\lambda)=f_{A}^{1}(\phi_{k},\lambda)
h_{A}^{1}(\phi_{k},\lambda)+f_{A}^{2}(\phi_{k},\lambda)
h_{A}^{2}(\phi_{k},\lambda)
\end{eqnarray}
After using equation (\ref{27}) and similar relations for
$h_{A}^{j_{A}}$ and $h_{A}^{j_{A}}$, we get the following
relations:
\begin{eqnarray}
f_{A}^{1}(\phi_{l},\lambda)
h_{A}^{1}(\phi_{l},\lambda)=f_{A}^{1}(\phi_{k},\lambda)
h_{A}^{1}(\phi_{k},\lambda)\\\nonumber f_{A}^{2}(\phi_{l},\lambda)
h_{A}^{2}(\phi_{l},\lambda)=f_{A}^{2}(\phi_{k},\lambda)
h_{A}^{2}(\phi_{k},\lambda)
\end{eqnarray}
That is appropriate result for derivation of eq. (\ref{13}) again.




\begin{thebibliography}{99}
\bibitem {EPR}  A. Einstein, B. Podolsky, and N. Rosen, Phys. Rev. $\mathbf{47}$, 777 (1935).

\bibitem {Bell} J.s. Bell, Physics (Long Island City, N.Y.) $\mathbf{1}$, 195 (1964); reprinted in Bell, J.S.
\emph{Speakable and Unspeakable in Quantum Mechanics} (Cambridge
Univ. Press, Cambridge, 1987).


\bibitem{As} A. Aspect, P. Grangier, and G. Roger,
\emph{Phys. Rev. Lett.} $\mathbf{49}$, 91-94 (1982); A. Aspect, J.
Dalibard, and G. Roger, \emph{Phys. Rev. Lett.} $\mathbf{49}$,
1804-1807 (1982); W. Tittel, \emph{et al.}, \emph{Phys. Rev.
Lett.} $\mathbf{81}$, 3563-3566 (1998); J. Pan, \emph{et al.},
\emph{ Nature} $\mathbf{403}$, 515 (2000); M. A. Rowe, \emph{et
al.}, \emph{ Nature} $\mathbf{409}$, 791 (2001); A. Vaziri, G.
Weihs, and A. Zeilinger, \emph{Phys. Rev. Lett.} $\mathbf{89}$,
240401 (2002); T. B. Pittman, and J. D. Franson, \emph{Phys. Rev.
Lett.} $\mathbf{90}$, 240401 (2003); Y. Huang, \emph{et al.}
\emph{Phys. Rev. Lett.} $\mathbf{90}$, 250401 (2003); Z. Zhao,
\emph{et al.}, \emph{ Phys. Rev. Lett. } $\mathbf{91}$, 180401
(2003); W. Tittel, and G. Weihs, \emph{ Q. Inf. Comp.}
$\mathbf{1}$, 3 (2001); Z. Chen, \emph{et al.}, Phys. Rev. Lett.
$\mathbf{90}$, 160408 (2003); C. Simon, C. Brukner, and A.
Zeilinger, Phys. Rev. Lett. $\mathbf{86}$, 4427 (2001); C. A.
Sackett, \emph{et al.}, Nature $\mathbf{404}$, 256 (2000).

\bibitem {Barre} J. Barrett, D. Collins, L. Hardy, A. Kent, and S. Popescu,
Phys. Rev. A $\mathbf{66}$, 42111 (2002).

\bibitem{As1} A. Aspect, \emph{ Nature} $\mathbf{398}$, 189-190
(1999); N. Gisin, and H. Zbinden, \emph{Phys. Lett. A}
$\mathbf{264}$, 103-107 (1999); P. G. Kwiat, \emph{et al.},
\emph{Phys. Rev. A} $\mathbf{49}$, 3209-3220 (1994); M.
Freyberger, \emph{et al.} \emph{Phys. Rev. A} $\mathbf{53}$,
1232-1244 (1996); A. Beige, W. J. Munro, and P. L. Knight,
\emph{Phys. Rev. A} $\mathbf{62}$, 0521021 (2000).

\bibitem {CHSH} J. F. Clauser, M. A. Horne, A. Shimony, and R. A. Holt, Phys. Rev. Lett.
$ \mathbf{23}$, 880 (1969).

\bibitem{CH} J. F. Clauser, and M.A. Horne, Phys. Rev.D $\mathbf{10}$, 526
(1974);

\bibitem{KS} S. Kochen, E. P. Specker, J. Math. $\mathbf{17}$, 59 (1967).



\bibitem {GHZ} D. M. Greenberger, M. A. Horne, A. Shimony, A. Zeilinger,
Am. J. Phys. $\mathbf{58}$, 1131 (1990).
\bibitem {Hardy} L. Hardy, Phys. Rev. Lett. $\mathbf{71}$, 1665 (1993).
\bibitem {Cab} A. Cabello, Phys. Rev. Lett. $\mathbf{86}$, 1911 (2001).


\bibitem {Gisin} V. Scarani, and N. Gisin, Phys. Lett. A $\mathbf{295}$, 167 (2002).

\bibitem{Zei} A. Zeilinger, \emph{ Nature} $\mathbf{408}$, 639-641 (2000);
A. Zeilinger, \emph{ Nature} $\mathbf{438}$, 743 (2005); Brukner
and A. Zeilinger, Phys. Rev. Lett. $\mathbf{83}$, 3354 (1999);
Brukner and A. Zeilinger, Phys. Rev. A $\mathbf{63}$, 022113
(2001); Brukner and A. Zeilinger, e-print arXiv:quantph/0212084;
Brukner and A. Zeilinger, Phil. Trans. R. Soc. Lond. A
$\mathbf{360}$, 1061 (2002), e-print arXiv:quant-ph/0201026; A.
Shafiee, F. Safinejad, F. Naqsh Foundations of Physics Letters,
$\mathbf{19}$, No. 1, (2006).





\bibitem{Wenger} J. Wenger, \emph{et al.},
Phys. Rev. A $\mathbf{67}$, 12105 (2003); H. Jeong, \emph{et al.},
Phys. Rev. A $\mathbf{67}$, 12106 (2003); Z. Chen, and Y. Zhang,
quant-ph/0103082; S. Massar, and S. Pironio, Phys. Rev. A
$\mathbf{64}$, 62108 (2001); P. van Loock, and S. L. Braunstein,
Phys. Rev. A $\mathbf{63}$, 22106 (2001), for compleat review read
S. L. Braunstein and P. Van Loock Rev. Mod. Phys. $\mathbf{77}$,
513 (2005).
\bibitem{Zuk} D. Kaszlikowski, and M. Zukowski, Phys. Rev. A $\mathbf{61}$,
22114 (2000).
\bibitem{Cab1} A. Cabello, Phys. Rev. A $\mathbf{63}$, 22104 (2001);
D. Kaszlikowski, \emph{et al.}, Phys. Rev. Lett. $\mathbf{85}$,
4418 (2000).
\bibitem{Cab2} A. Cabello, Phys. Rev. Lett.
$\mathbf{88}$, 60403 (2002).













\bibitem{Brass} G. Brassard, R. Cleve and A. Tapp, Phys. Rev. Lett. $\mathbf{83}$,
1874 (1999); G. Brassard, quant-ph/0101005.

\bibitem{Bacon} B. F. Toner and D. Bacon, Phys. Rev. Lett. $\mathbf{91}$, 187904 (2003);
D. Bacon and B. F. Toner , Phys. Rev. Lett. $\mathbf{90}$, 157904
(2003); K. Svozil, Phys. Rev. A \textbf{72} 050302 (R) (2005).

\bibitem{Cerf} S. Popescu and D. Rohrlich, Found. Phys. $\mathbf{24}$, 379 (1994);
N. J. Cerf, N. Gisin, S. Massar, and S. Popescu, Phys. Rev. Lett.
$\mathbf{94}$, 220403 (2005); J. Barrett and S. Pironio,  Physical
Review Letters $\mathbf{95}$ 140401 (2005); N. Brunner, N. Gisin
and V. Scarani, New Journal of Physics $\mathbf{7}$ 88 (2005).

\bibitem {Short} A. J. Short \emph{et al.}, Phys. Rev. A $\mathbf{73}$, 012101
(2006).

\bibitem{BP} Barrett, J. and Pironio, S., (2005)  Phys. Rev. Lett. $\mathbf{95}$
140401.

\bibitem {Fa} A. Fahmi, and M. Golshani, Phys. Lett. A,
$\mathbf{306}$, 259, (2003).

\bibitem{Bohm1} For example in ref \cite{Bohm}, p. 140, the authors have defined some sets of hidden
variables where the result of measurement depend on them (hidden
variables $\mu_{a}$ and $\mu_{b}$ associated with the
corresponding pieces of measuring apparatus of the first and
second particles respectively. In addition there will be some sets
of variables $\lambda_{A}$ and $\lambda_{B}$, belonging to the
particle $A$ and $B$ themselves).

\bibitem{Bohm} D. Bohm, and B. J. Hiley, \emph{The Undivided Universe: an ontological
interpretation of quantum theory} (Routledge 1993).



\bibitem{Hardy1} L. Hardy, Phys. Lett. A $\mathbf{161}$, 21
(1991).








\bibitem{Peres} A. Peres, private communication. A. Peres, and
D. R. Terno, \emph{Rev. Mod. Phys.} $\mathbf{76}$, (2004).

\bibitem{Brass1} G. Brassard, \emph{ Nature} $\mathbf{1}$, 2-4 (2005); G. Brassard, \emph{et al.}, Phys. Rev. Lett. $\mathbf{96}$,
250401 (2006).












































\end{thebibliography}
\end{document}